\documentclass[english]{article}
\usepackage[latin9]{inputenc}
\usepackage{color}

\newcommand{\lyxaddress}[1]{
\par {\raggedright #1
\vspace{1.4em}
\noindent\par}
}

\usepackage{babel}

\begin{document}

\title{\textbf{Presentation of the }\textbf{\emph{Second Big Challenge Symposium
- The Big Challenge of Cosmological Understanding: Gravitation, Dark
Matter and Dark Energy. Towards New Scenarios}}}

\author{\textbf{Christian Corda }}

\maketitle

\lyxaddress{\begin{center}
Associazione Scientifica Galileo Galilei, Via Bruno Buozzi 47 - 59100
PRATO, Italy 
\par\end{center}}

\begin{center}
\textit{E-mail address:} \textcolor{blue}{cordac.galilei@gmail.com}
\par\end{center}
\begin{itemize}
\item \textbf{This Symposium is devoted to the Memory of Lev Kofman June-17-1957-November-12-2009}
\end{itemize}
The accelerated expansion of the universe that is today observed suggests
that cosmological dynamics is dominated by the so-called Dark Energy
field which provides a large negative pressure. This is the standard
picture, in which such new ingredient is considered as a source of
the \textit{right hand side} of the field equations. It should be
some form of non-clustered non-zero vacuum energy which, together
with the clustered Dark Matter, drives the global dynamics. This is
the so-called {}``\emph{concordance model}'' ($\Lambda$CDM) which
gives, in agreement with the Cosmic Microwave Background Radiation
(CMBR), dim Lyman Limit Systems (LLS) and type la supernovae (SNeIa)
data, a good framework to understand the today observed Universe.
However, it presents several shortcomings as the well known ''\emph{coincidence}''
and \emph{{}``cosmological constant}'' problems \cite{key-1}. An
alternative approach is to change the \emph{left hand side} of the
field equations, and check if observed cosmic dynamics can be achieved
by extending general relativity \cite{key-2,key-3,key-4,key-5,key-6}.
In this different context, it is not required to search candidates
for Dark Energy and Dark Matter, which till now have not been found.
Rather, one can only stand on the \emph{{}``observed}'' ingredients:
curvature and baryon matter, to account for the observations. Considering
this point of view, one can think that gravity is not scale-invariant.
Such an assumption opens a room for alternative theories to be introduced
\cite{key-7,key-8,key-9}. In principle, the most popular Dark Energy
and Dark Matter models can be achieved by considering $f(R)$ theories
of gravity \cite{key-2}-\cite{key-9}, where $R$ is the Ricci curvature
scalar. 

In this picture, even the sensitive detectors for gravitational waves
like bars and interferometers (i.e. those which are currently in operation
and the ones which are in a phase of planning and proposal stages)
\cite{key-10} could, in principle, be important to confirm or rule
out the physical consistency of general relativity or of any other
theory of gravitation. This is because, in the context of Extended
Theories of Gravity, some differences between General Relativity and
the alternative theories can be pointed out as far as the linearized
theory of gravity is concerned \cite{key-11,key-12,key-13}.

In particular, it has been recently shown that the interferometric
detection of gravitational waves will be the definitive test for General
Relativity or alternatively a strong endorsement for Extended Theories
of Gravity \cite{key-13}.

The goal of this Symposium is to obtain a tapestry of the present
status of theory and observations concerning Gravitation and Dark
Universe.

\subsection*{Acknowledgements}

The Associazione Scientifica Galileo Galilei and the Institute for
Basic Research have to be thanked for supporting this proceeding and
the whole \emph{Second Big Challenge Symposium}.

\end{document}